\documentclass[aps,pre,
superscriptaddress,groupedaddress,showpacs,twocolumn]{revtex4}
\usepackage{graphicx,epsf}
\usepackage{hyperref}
\usepackage{xcolor}
\usepackage{psfrag}
\usepackage{amsmath}
\usepackage{amssymb}
\usepackage{orcidlink}
\usepackage{psfrag}
\usepackage[english]{babel}
\usepackage[title]{appendix}
\begin{document}

\title{Mapping self-avoiding walk on {obstacle-ridden} lattice 
onto chelation of heavy metal ions: Monte Carlo study}


\author{Viktoria Blavatska\orcidlink{0000-0001-6158-1636}}
\affiliation{Institute for Condensed
Matter Physics of the National Academy of Sciences of Ukraine,\\
1 Svientsitskii Str., UA-79011 Lviv, Ukraine}
\email{blavatskav@gmail.com}
\author{Jaroslav Ilnytskyi\orcidlink{0000-0002-1868-5648}}
\affiliation{Institute for Condensed
Matter Physics of the National Academy of Sciences of Ukraine,\\
1 Svientsitskii Str., UA-79011 Lviv, Ukraine}
\affiliation{Institute of Applied Mathematics and Fundamental Sciences, \\
Lviv Polytechnic National University, 12 S. Bandera Str., UA-79013 Lviv, Ukraine}
\author{Erkki L\"ahderanta\orcidlink{0000-0002-1596-2849}}
\affiliation{Department of Physics, School of Engineering Science,
LUT University, Yliopistonkatu 34, FI-53850 Lappeenranta, Finland}
\affiliation{Department of Physics, Universitat de les Illes Balears, Cra Valldemossa, km. 7.5, 07122, Palma, Spain}

\begin{abstract}


Self-avoiding walk (SAW) represents linear polymer chain on a large scale, neglecting its chemical details and emphasizing the role of its conformational statistics. The role of the latter is important in formation of agglomerates and complexes involving polymers and organic or inorganic particles, such as polymer-stabilized colloidal suspensions, microemulsions, or micellar solutions. When such particles can be adsorbed on a polymer of considerably larger dimensions than themselves, this setup may represent chelation of heavy metal ions by polymeric chelants. We consider the SAW of the length $N$ on a cubic lattice ridden by randomly distributed obstacles of the concentration $p$ interpreted as ions. The SAW monomers can bind to the obstacles with variable binding energy $\varepsilon$ mimicking formation of the chelation bond. Pruned-enriched Rosenbluth method (PERM) Monte Carlo (MC) algorithm is applied to simulate system behaviour. {We focus on several relevant properties related to the chelation efficiency and strength, as functions of the variables set $\{p,N,\varepsilon\}$. The results are interpreted in terms of conformational freedom, excluded volume effects and loop formation for the SAW, and the tendencies being predicted are in agreement with some experimental data.}
\end{abstract}
\date{\today}
\maketitle


\section{Introduction}

When polymers mix with organic or inorganic particles, a whole new range of composite systems emerge \cite{Hanemann2010, Toyoda2019}. These range from the polymer-stabilized colloidal suspensions, microemulsions, micellar solutions \cite{Dunn1986, Ottewill1990, Cabane1997, Shipway2000, Rensmo2002}, to smart materials \cite{Bahl2020, Jain2023}, where the specific features of particles embedded into a polymer matrix define the response of such material on external stimuli. Typically, two limit cases are distinguished depending on the relation between the characteristic dimensions of a polymer, given by its radius of gyration $R_g$, and the dimension of a particle, e.g. its diameter $D$ if the particle has a spherical shape. If $R_g\sim D$, then polymers can act either as spacers and separate particles in colloidal dispersion, or bind them together in the case of adsorption of polymers on their surface \cite{Lafuma1991} due to adhesion of their monomers. The latter occurs owing to covalent bonding, van der Waals, or electrostatic interactions. The conditions for adsorption involve both the particle diameter $D$, and the adhesion energy per a monomer, as shown in an early theoretical study by Pincus et al. \cite{Pincus1984}. In the opposite case, $R_g > D$, several scenarios are possible, depending on polymer concentration and the adsorbing ability of particles. Cabane et al. \cite{Cabane1997} distinguished some of them as: depletion flocculation for the case of  nonadsorbing polymers; bumper effect for adsorbing polymers forming crowded polymer layers around each particle; and sticker effect, when adsorbing polymers bridge particles together. The latter scenario results in formation of necklace-like aggregates when several particles are wrapped over by a single polymer chain \cite{Wong1992,Spalla1993}.

The case of $R_g > D$ is also directly related to adsorption of dissolved polymer on small size colloidal particles {\cite{Becher1990, Gonzlez2011, Fortuna2015, GonzlezGarca2020}, with the emphasis often put on the efficiency of this process, e.g. its dependence on the polymer length \cite{Gonzlez2011}}. In this context, a number of theoretical studies exist. A microscopic theory based on the scaling approach has been developed \cite{Klimov1992, Marques1988, Aubouy1998}, resulting in the effective free energy estimates 
for the polymer coating obtained via integral RISM equation technique \cite{Khalatur1997}. MC simulations of the polymer-colloid systems have been also carried out in Refs.~\cite{Zherenkova1998, Khalatur1999}. An estimation of the maximum number $N_a^{\mathrm{max}}$ of particles that can be adsorbed on a single polymer molecule, 
provided that the size of particles $a$ is comparable with the statistical segment length, was found to scale as $N_a^{\mathrm{max}}\sim (R_g/a)^3$ \cite{Nowicki2002}. Computer simulations of the SAW in a presence of small colloidal particles have been also reported in Ref.~\cite{Nowicki2002}, where, in particular, the conformational entropies and segment distributions in the coils perturbed by presence of particles of various size and shape were calculated. 

One can envisage the extreme case, $D \ll R_g$, when $D$ is on a atomic scale, and this case can be related to modelling chelation of heavy metal ions by polymeric chelants. This problem is of a paramount importance from both biomedical aspect related to deintoxication {therapy} \cite{Bernhoft2012, Paduraru2022, Genchi2020, Prasad2021}, and environmental perspective of wastewater cleaning \cite{Singh2018, Qasem2021}. {Typical scenario is considered in Refs.~\cite{Rosthauser1981, Johann2019, Tyagi2020}, where the polymeric scaffold contains chelating groups and both intra- and inter-chain complexation with heavy metal ions are possible.}

{The process of chelation is intrinsically multi-scale, since the events occuring at both atomic and macroscopic scales affect its efficiency. Atomic scale provides conditions for formation and stability of chelation complexes between specific chelants and particular metal ions. There is a range of widely used chelants such as: ethylenediaminetetraacetic acid (EDTA), 2,3-dimercaptopropanesulfonic acid, 2-3 dimercaptosuccinic acid, polyetheleneimine, etc. \cite{Jia2014, Zhang2019, Kuo2001, Nam2015}, as well as some others developed quite recently \cite{Miller2014, BoehmSturm2018}.}

{Macroscopic scale of a problem is equally important because of a high molecular weight of involved polymeric chains.} Here the statistics of macromolecular conformations of a polymer plays the predominant role, and it can be attained by considering the SAW model on a lattice \cite{deGennes1979, desCloizeaux1982, Freed1981, Hooper2020}. { Despite its simplicity, the SAW model proved to be very efficient in capturing the universal conformational properties of polymers in good solvent regime. In particular, it was successfully applied in describing the coil-globule transitions \cite{Bedini2013}, polymer behaviour in confined space \cite{Laachi2010,Benito2018,Bradly2022}, adsorption processes of polymers on disordered structures \cite{Blavatska2012,Bubanja1993}, etc.}

Within this modelling approach, heavy metal ions can be represented as obstacles occupying some of free sites of the lattice, whereas formation of chelation complexes -- by effective attraction between them and the monomers of the SAW. One should remark, that intrinsically dynamic process of chelation, where both a polymeric chelant and an ion are diffusing throughout the system, can be mapped onto the ensemble averaging philosophy of the MC method. In particular, one can replace the time trajectory for moving obstacles by an ensemble of discrete ``snapshots'', each with frozen random distribution of obstacles throughout the system. The SAW is grown on every such realization, and the results are averaged over the ensemble, thus taking into account motion of obstacles implicitly. 
To the best of our knowledge, such connection between the SAW in the environment containing impurities and chelation of heavy metal ions in the context of water cleaning, has not been exploited before.


The goal of a current study is to map the SAW {on an obstacle-ridden lattice onto the problem of chelation of heavy metal ions. We analyze the effect of polymer length, obstacles concentration, and the magnitude of a binding energy between the SAW and obstacles, on a set of properties related to the efficiency and strength of chelation.} The outline of a study is as follows. Sec.~\ref{sec:1} contains description of the approach used in this study, in Sec.~\ref{sec:2} we present and discuss obtained results, the study is finalised by Conclusions.

\begin{figure}[b!]
 \begin{center}
 \includegraphics[width=10.2cm]{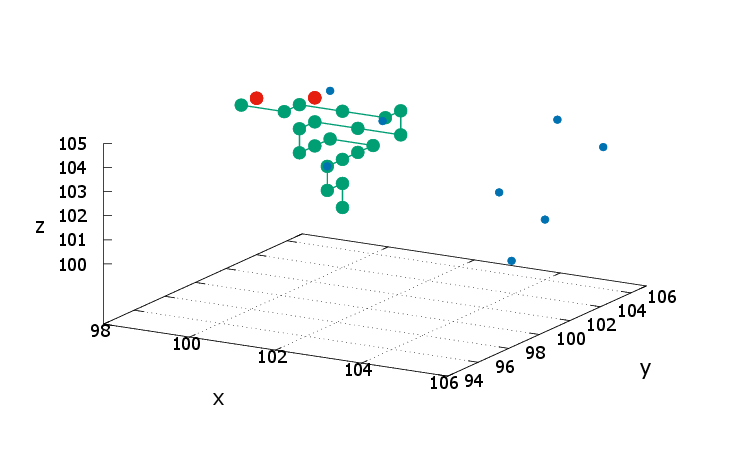}
\includegraphics[width=10.2cm]{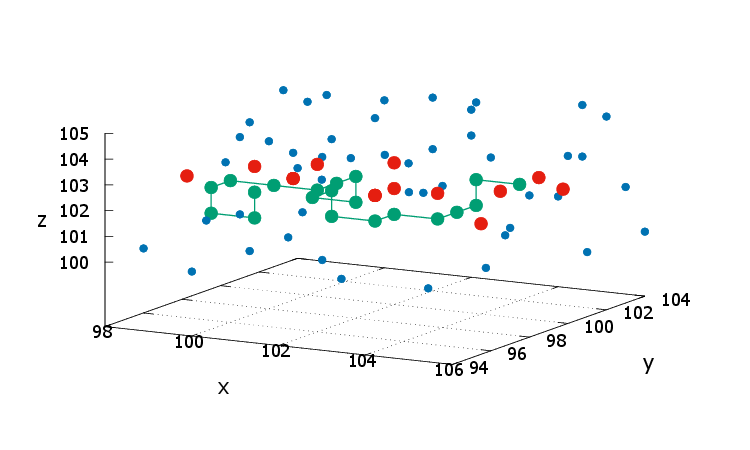}
\end{center}
\caption{\label{SAWcheme} (color online) A snapshot of SAW trajectory of $N=20$ monomers (shown by green symbols) on a lattice with fraction $p=0.01$ (upper panel) and $p=0.1$ (lower panel) of point-like obstacles shown by blue symbols. Obstacles absorbed to SAW trajectory are shown in red. Only the part of lattice around the SAW trajectory is shown.}
\end{figure}

\section{The method}\label{sec:1}

We start with the $3D$ simple cubic lattice of $L^3$ sites with the linear dimension of $L = 400$. Initial configuration contains the fraction $p$ of immobilized obstacles distributed at random positions, each interpreted as either a small size colloid or heavy metal ion. The range of $0.01<p<0.1$ is considered, and the results are averaged over 1000 replicas with different spatial arrangement of obstacles.

We prescribe a label  $r(x(i), y(i), z(i))$ to each $i$th site of the lattice with coordinates $\{x(i),y(i),z(i)\}$ such that 
$$
r(x(i), y(i), z(i)){=}\begin{cases}
1, \, \,   \text{if $i$th site contains an obstacle,}\\
0, \, \,   \text{otherwise. }
\end{cases} 
$$
To construct SAW on such obstacle-ridden lattice, we apply the pruned-enriched Rosenbluth Method (PERM) \cite{Grassberger97}, based on the  Rosenbluth-Rosenbluth (RR) method originated from early 50-ies \cite{Rosenbluth55}  and enrichment strategies \cite{Wall59}. The polymer trajectory is grown step by step, i.e., the $n$th monomer is placed at a randomly chosen unoccupied site, that is the nearest neighbour to the previous, $(n-1)$th, monomer, until the required polymer length, $N$, is achieved (as presented in Fig. \ref{SAWcheme}). 
Let $\vec{R}_n=\{ X_n, Y_n, Z_n \}$ define the spatial position of $n$th monomer of the chain. A short-range attraction between each neighbouring monomer and an obstacle, characterized by a binding energy $\varepsilon$, provides formation of the chelation bond. The total binding energy of the $n$-monomer chain is  then given by
\begin{equation}
E_n=N_s(n)\varepsilon,
\label{energy}
\end{equation}
where $N_s$ is the number of nearest neighbour monomer-obstacle contacts for a given realization of the SAW trajectory, which can be given by 
\begin{eqnarray}
&&N_s(n)=\sum_{l=1}^n \Big(r(X_l\pm1,Y_l,Z_l)+\nonumber\\
&&+r(X_l,Y_l\pm1,Z_l)+r(X_l,Y_l,Z_l\pm1)\Big). \label{Nsdef}
\end{eqnarray}

A weight $W_n$ is prescribed to each  configuration at the $n$th step, given by the combination of the Rosenbluth weights \cite{Rosenbluth55}
\begin{equation}
W_n^{{\rm Rosenbluth}}= \prod_{l{=}2}^n m_l,
\end{equation}
where  $m_l$ is the number of available free lattice sites for adding the $l$th monomer, as well as the respective Boltzmann probability \cite{Bachmann04}
\begin{equation}
W_n= W_n^{{\rm Rosenbluth}} \exp\left[\frac{-(E_n-E_{n-1})}{k_B T}\right].
\label{weight}
\end{equation}
Here, $k_B$ is the Boltzmann {factor}, $T$ is the temperature. In what follows, we will assume the units in which $k_BT=1$, thus measuring $\varepsilon$ in units of $k_BT$.

Population control in PERM suggests {\it pruning} configurations with too small weights, and {\it enriching} the sample with copies of the high-weight configurations \cite{Grassberger97}.  Pruning and enrichment are performed by choosing thresholds $W_n^{<}$ and $W_n^{>}$ depending on the estimate of the partition sum $Z_n=\sum_{{\rm conf}} W_n^{{\rm conf}}$ of the $n$-monomer chain, where summation is performed over all available configurations of a chain. If the current weight $W_n$ of an $n$-monomer chain is less than $W_n^{<}$, a random number ${\rm rand} {=}\{0,1\}$ is chosen; if ${\rm rand}{=}0$, the chain is discarded, otherwise it is kept and its weight is doubled. Thus, low-weight chains are pruned with probability $1/2$. If $W_n$ exceeds  $W_n^{>}$, the configuration is doubled and the weight of each copy is taken as half the original weight. We adjust the pruning-enrichment control parameter such that on average 10 chains of total length $N$ are generated per each iteration \cite{Bachmann04}, and perform $10^6$ iterations. 

Each iteration for a chain of total length $N$ starts from the same initial point, until the desired number of chain configurations is obtained. The thermodynamic averaging for any observable of interest $O$ is given by
\begin{equation}
 \overline {O } =\frac{{\sum_{{\rm conf}} W_N^{{\rm conf}} O }} {Z_N} \label{confaver}
 \end{equation}
 with  $W_N^{{\rm conf}}$ being the weight of an $N$-monomer chain in a given configuration given by (\ref{weight}). The disorder averaging \begin{equation}
  \langle {\overline {O}} \rangle= \frac{1}{M}\sum_{i=1}^M {\overline {O}}^i \label{aver}
 \end{equation}
 is  performed over $M=1000$ different realizations of random spatial arrangement of obstacles, where ${\overline{O}}^i$ is thermodynamic averaging in $i$th replica of a lattice. 

\section{Results}\label{sec:2}

\begin{figure}[b!]
 \begin{center}
\includegraphics[width=8.2cm]{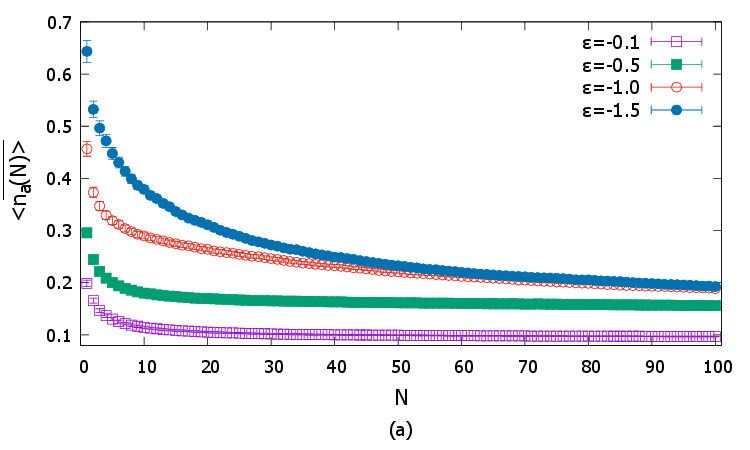}
\includegraphics[width=8.2cm]{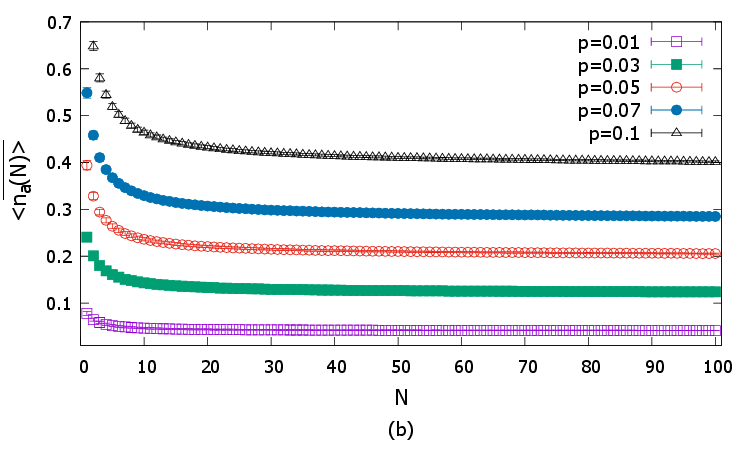}
\includegraphics[width=8.2cm]{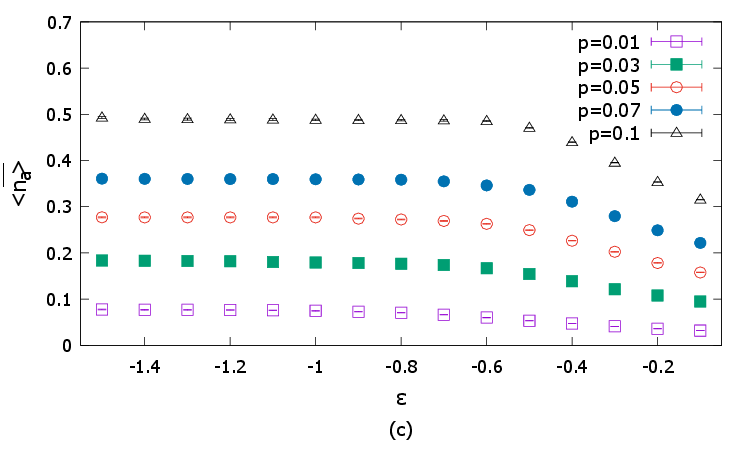}
\includegraphics[width=8.2cm]{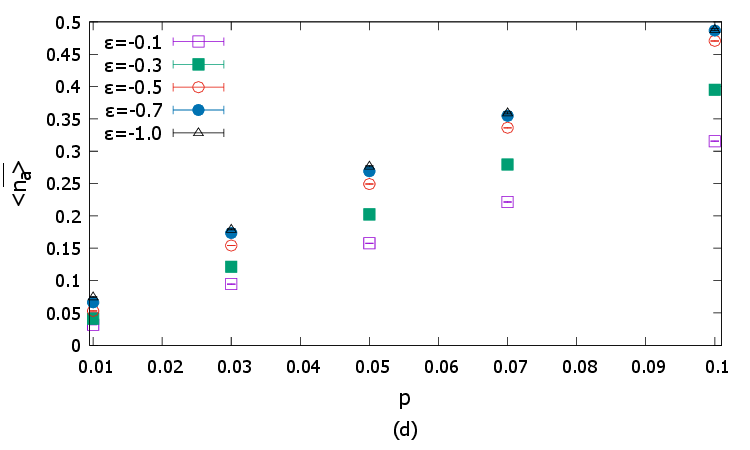}
\end{center}
\caption{\label{Na} Adsorption efficiency as given by the average number of adsorbed obstacles normalized to the length of chain $\langle \overline{n_a(N)} \rangle$ a)  at fixed $p=0.03$ and several values of  $\varepsilon$ and b) at fixed $\varepsilon=-0.3$ and several values of  $p$.
Asymptotic values of  $\langle \overline{n}_a \rangle$ {at $N\to\infty$}, obtained on the basis of fitting function (\ref{as}), as functions of $\varepsilon$ c) and function of $p$ d). 
}
\end{figure}

We start with evaluation of the average number $N_a$ of obstacles, that are adsorbed by a polymer chain of a length $N$. 
To this end, we prescribe labels  $v(x(i),y(i),z(i))$ to each $i$th site of the lattice containing an obstacle, with initial values $v(x(i),y(i),z(i))=0$.  
When constructing a single SAW trajectory, at each  $n$th step, we are checking the conditions:
 \begin{itemize}
     \item 
 if $ r(X_n\pm1,Y_n,Z_n)=1$, then 
$ v(X_n\pm1,Y_n,Z_n)=v(X_n\pm1,Y_n,Z_n)+1$;
     \item 
 if $ r(X_n,Y_n\pm1,Z_n)=1$, then 
$ v(X_n,Y_n\pm1,Z_n)=v(X_n,Y_n\pm1,Z_n)+1$;
 \item 
 if $ r(X_n,Y_n,Z_n\pm1)=1$, then 
$ v(X_n,Y_n,Z_n\pm1)=v(X_n,Y_n,Z_n\pm1)+1$;
 \end{itemize}
 so that at the end the values of $v(x(i),y(i),z(i))$ contain the total number of bonds of an obstacle positioned at site $i$ with monomers of SAW strajectory. Thus, the number of different obstacles  $N_a$ encountered by  SAW trajectory  (as visualised on Fig. \ref{SAWcheme}) is given by
 \begin{eqnarray}
&&N_a(n)=\sum_{l=1}^n \Big(\delta_{v(X_l\pm1,Y_l,Z_l),1}+\nonumber\\
&&+\delta_{v(X_l,Y_l\pm1,Z_l),1}+\delta_{v(X_l,Y_l,Z_l\pm1),1}\Big),
\end{eqnarray}
where $\delta$ is the Kronecker delta, i.e. the summation is carried on by those obstacles, which have been encountered only once.
The thermodynamic and disorder averaged value of $\langle N_a \rangle $ is thus obtained according to Eqs. (\ref{confaver}) and (\ref{aver}).

\begin{figure}[b!]
 \begin{center}
\includegraphics[width=8.2cm]{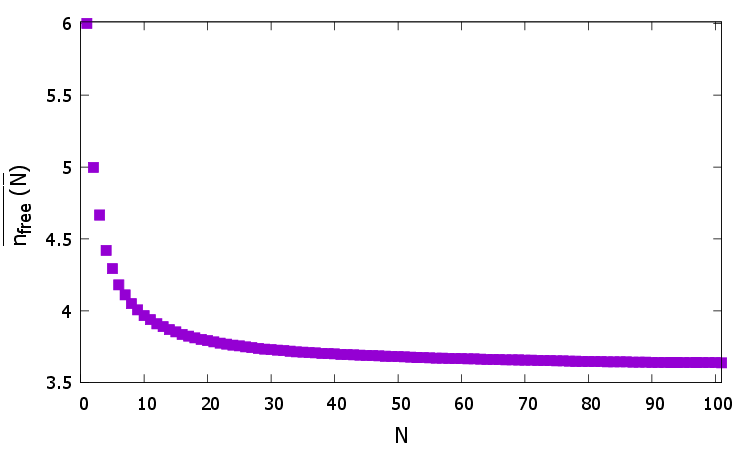}
\end{center}
{\caption{\label{nfree} Averaged number of ``free'' nearest neighbour sites of SAW trajectory per one monomer as function of $N$.}}
\end{figure}
The normalized value, $n_a(N)=N_a(N)/N$, is associated with adsorption efficiency for a polymer chain of the length $N$.
{ It is the function of the chain length, $N$, binding energy, $\varepsilon$, and obstacles concentration, $p$. Obtained results are collected in Fig.~\ref{Na}a-c.}

Dependence of an average adsorption efficiency, $\langle \overline{n_a(N)} \rangle$, on $N$, evaluated at a range of binding energies $\varepsilon$ and at fixed obstacles concentration, $p=0.03$, is shown in Fig.~\ref{Na}a. Similar behavior is found at other concentrations, $p<0.1$. The first result to point out is, that at each given $\varepsilon$, $\langle \overline{n_a(N)} \rangle$ decreases with $N$. Therefore, shorter chains are found to be more efficient in adsorbing obstacles then longer ones, at the same binding energy $\varepsilon$. {This result recalls some experimental findings by G\'{o}nzalez et al. for the flocculation of colloidal silica particles by blends of the EDTA-containing polyelectrolytes \cite{Gonzlez2011}. By using a blend of short and long polymers in various proportions, the authors demonstrated that the presence of short polymers improves the fraction of sedimented particles. The effect, however, could be partly associated with the efficiency of the surface charge neutralization of the particles.} The second result is that, the increase of the magnitude of binding energy, $|\varepsilon|$, raises adsorption efficiency of shorter chains to much higher extent than in the case of longer chains.

We can explain these two results by the suggestion, that, for the case of short chains, adsorption is driven mainly by the magnitude of binding energy, whereas, for longer chains, it is {restricted by a self-avoidance of chains originated from the excluded volume effects. This factor was also pointed out by Fortuna et al. \cite{Fortuna2015} in their simulation of chelating effect in polymeric binders with increased affinity and selectivity. To go into this effect in more depth, we considered the total number $N_{{\rm free}}(N)$  of  ``free” neighbor sites around each monomer of SAW of length $N$, which are not occupied by SAW trajectory. E.g., for $N=1$, one obviously has $N_{{\rm free}}(1)=6$, whereas for $N=2$, due to the bond between two monomers, one has $N_{{\rm free}}(2)=10$. The averaged normalized value ${\overline{n_{{\rm free}}(N)}}={\overline{N_{{\rm free}}(N)}}/N$ per one monomer is presented in Fig. \ref{nfree}. It decreases with $N$ and gradually reaches the asymptotic value equal to $z(3)-1$, where $z(3)=4.68404(9)$ \cite{MacDonald2000} is the so-called fugacity of SAW.}

{ Dependence of $\langle \overline{n_a(N)} \rangle$ on $N$ at fixed $\varepsilon=-0.3$ and at several values of obstacle concentrations, $p$, is shown in Fig.~\ref{Na}d. It indicates monotoneous growth with $p$, as expected (see also Fig. \ref{SAWcheme}), and at the concentration range of $p\leq 0.1$, this dependence is close to linear, $\langle \overline{n_a(N)}\rangle \sim p$ for $\varepsilon=-0.3$ displayed here.}

{At large $N$, $\langle \overline{n_a(N)} \rangle$ saturates and it makes sense to find the asymptotic value, $\langle \overline{n_a} \rangle = \lim_{N\to\infty} \langle n_a(N) \rangle$, from the finite-size scaling law }
\begin{equation}
\langle {\overline{n_a(N)}} \rangle= 
 \langle \overline{n_a} \rangle+A/N, \label{as}
\end{equation}
which involves a constant $A$, and the correction to scaling terms have been neglected. To find the region of applicability of this expression, we estimate the lower cutoff, $N_{\mathrm{min}}$, such as at $N>N_{\mathrm{min}}$, the correction to scaling terms vanish. The $\chi^2$ value (sum of squares of normalized deviation from the regression line) divided by the number of degrees of freedom, $DF$ serves as an estimate for the fit accuracy. An example is given in Table \ref{fit}. 
\begin{table}[t!]
\begin{center}
    \begin{tabular}{|c|c|c|c|}
    \hline
        $N_{\mathrm{min}}$ & $\langle \overline{n_a}\rangle$ & $A$ &  $\chi^2/DF $ \\ \hline
       2 &  0.156(1) & 0.258(6) & 1.1280   \\ \hline	
       5 &  0.155(1)   & 0.265(5)  & 1.0001  \\ \hline	
       10  &  0.1545(8)  &  0.269(5) &  0.7753 \\ \hline         15  &  0.1547(5) &  0.270(5)   & 0.5985 \\ \hline
            \end{tabular}
    \caption{Results of the least-square fits for evaluation the asymptotic value $\langle \overline{n_a}\rangle$ when varying the lower cutoff $N_{\mathrm{min}}$ for the case $p=0.3$, $\varepsilon=-0.5$.  }
    \label{fit}
    \end{center}
\end{table}

{ The estimates for the asymptotic value, $\langle \overline{n_a} \rangle$, as the function of $\varepsilon$ at various $p$, are given in Fig. \ref{Na}c. At each fixed concentration $p$, $\langle \overline{n_a} \rangle$ raises upon the increase of the binding energy magnitude $|\varepsilon|$ saturating at some $p$-dependent value. The latter is close to $\langle \overline{n_a} \rangle \approx 5p$, following linear dependence of $\langle \overline{n_a}(N) \rangle$ on $N$, see Fig. \ref{Na}b.} Note that the similar behaviour with the threshold value of $\varepsilon$ around $0.5$ was observed while analyzing the colloid-particle contacts in Ref. \cite{Khalatur1999}.

{ Another property of interest is the average number of bonds $\langle \overline{n_{{\rm bond}}(N)} \rangle$ (``bridge points'') per single adsorbed obstacle, which is associated with the adsorption strength of a polymer with the length $N$. Indeed, when adsorbed with more bonds, an obstacle will have less probability to be desorbed by thermal fluctuations or a flow. Note, that because of a coarse-grained nature of the model, each bond does not represent chelation bond in a chemical sense, but rather a whole chelation complex.}
\begin{figure}[t!]
 \begin{center}
\includegraphics[width=8.2cm]{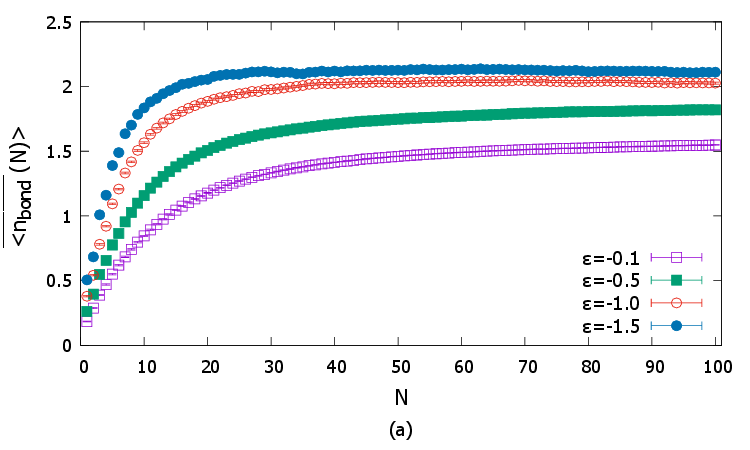}
\includegraphics[width=8.2cm]{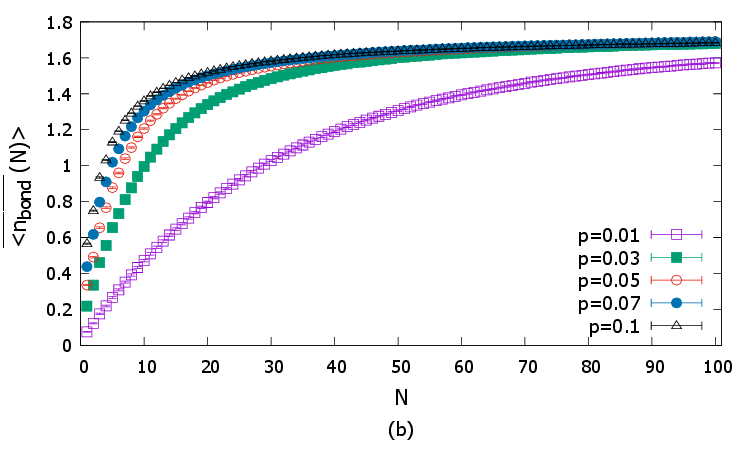}
\includegraphics[width=8.2cm]{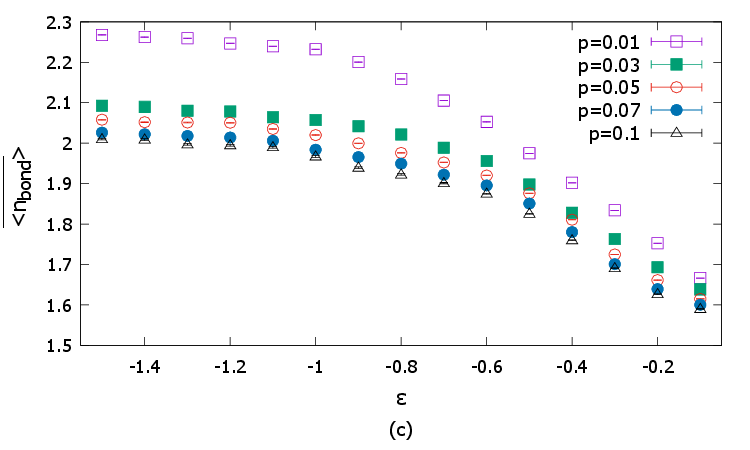}
\includegraphics[width=8.2cm]{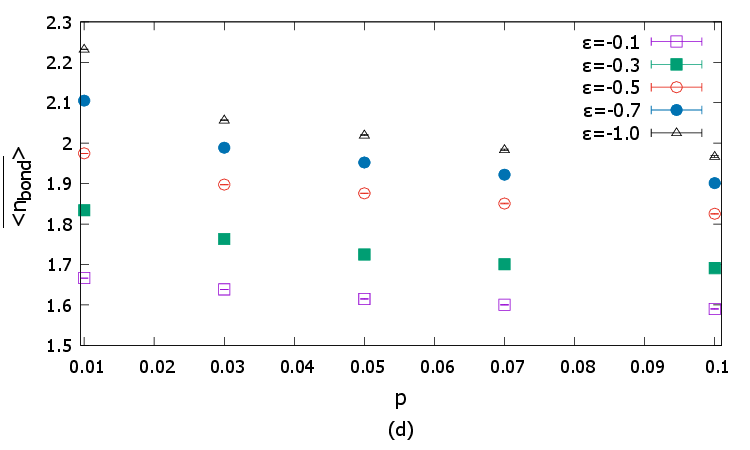}
\end{center}
\caption{\label{nbond} Adsorption strength as given by the average number $\langle {\overline{n_{\rm bond}(N)}} \rangle $ of bonds between a single obstacle particle and polymer chain as function of $N$ a) at fixed $p=0.03$ and several values of  $\varepsilon$ and b) at fixed $\varepsilon=0.3$ and several values of  $p$.  Asymptotic values of  $\langle {\overline {n_{{\rm bond}}}} \rangle$, obtained on the basis of fitting function (\ref{bs}), as functions of $\varepsilon$ c) and function of $p$ d)
}
\end{figure}

{ The results for adsorption strength, $\langle \overline{n_{{\rm bond}}(N)} \rangle$, performed at $p=0.03$ and several fixed values of $\varepsilon$, are presented in Fig. \ref{nbond}a. We found its dependence on $N$ to be an opposite to that for the adsorption efficiency, $\langle \overline{n_a(N)}\rangle$, Fig. \ref{Na}a. Low values for $\langle \overline{n_{{\rm bond}}(N)} \rangle$, observed for short chains, are the consequence of their limited possibilities to bend, and, hence, inability to form more than one bond with the same obstacle. For long chains, the conformation space is enriched by bend and looped shapes, and a longer chain can form multiple bonds with a single obstacle, thus increasing $\langle \overline{n_{{\rm bond}}(N)} \rangle$. We also note, that $\langle \overline{n_{{\rm bond}}(N)} \rangle$ is affected by $\varepsilon$ more strongly for the case of short chains comparing to the long ones, similarly to $\langle \overline{n_a(N)} \rangle$}.

\begin{figure}[b!]
 \begin{center}
\includegraphics[width=8.2cm]{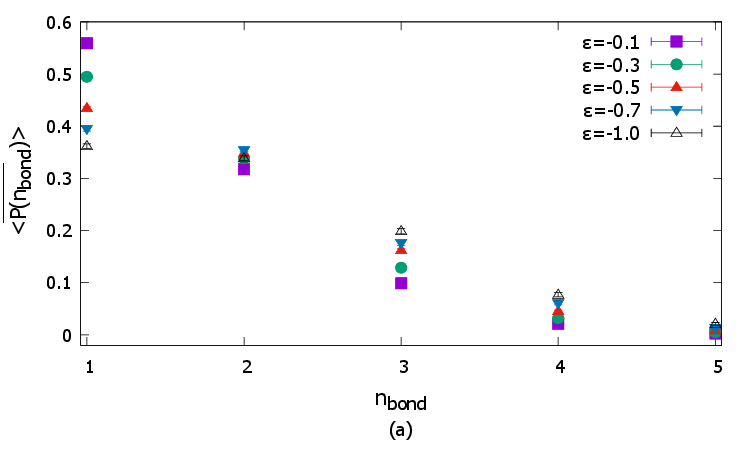}
\includegraphics[width=8.2cm]{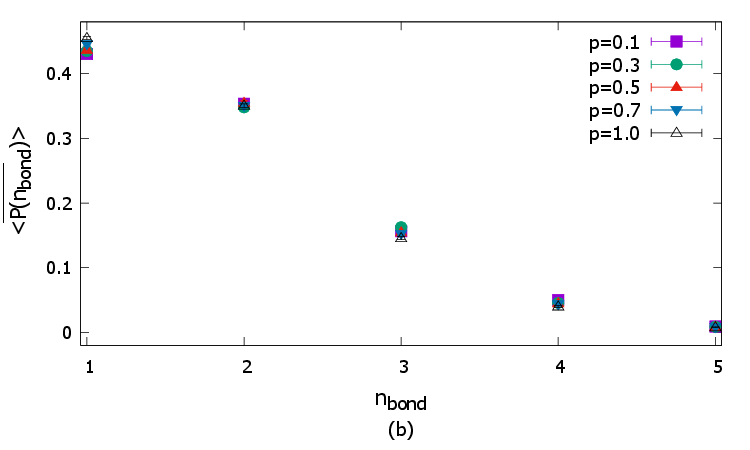}
\end{center}
{\caption{\label{probbond} Probability distribution of the number of bond per single adsorbed particle at fixed $N=100$. a) Fixed $p=0.03$ and a range of values for $\varepsilon$; b) fixed $\varepsilon=-0.5$ and a set of values of $p$.}}
\end{figure}

{ We also examined the probability distribution $P(n_{\rm bond})$ for the number of bonds, $n_{\rm bond}$, between a single adsorbed obstacle and a SAW. The results for the longest SAW, $N=100$, at fixed obstacles concentration, $p=0.03$, are shown in Fig. \ref{probbond}. Our first remark is that, the probability $P(n_{\rm bond})$ monotonically decreases with the increase of $n_{\rm bond}$ for all binding energies $\varepsilon$. This indicates that a small number of bonds always prevail. The shape of this decay, however, is affected by the value of $\varepsilon$. In particular, $P(1)$ is the highest for the weakest binding, $\varepsilon=-0.1$, whereas $P(n_{\rm bond})$ at $n_{\rm bond}\geq 2$ is the highest for the strongest adsorption case, $\varepsilon=-1.0$. This shows that strengthening of the binding between an obstacle and a SAW increases occurence of multiple bonds, $n_{\rm bond}\geq 2$. This explains the increase of $n_{{\rm bond}}$ at larger magnitudes of a binding energy, see Fig.~\ref{nbond}. Similar scenario is found for other lengths of a SAW, $N$. We found rather weak, if any, dependence of $P(n_{\rm bond})$ on the obstacles concentration, $p$, at fixed binding energy. The case of $\varepsilon=-0.5$ is demonstrated in Fig.~\ref{probbond}(b), whereas similar behavior is found at other values of $\varepsilon$.}

\begin{figure}[t!]
 \begin{center}
\includegraphics[width=8.2cm]{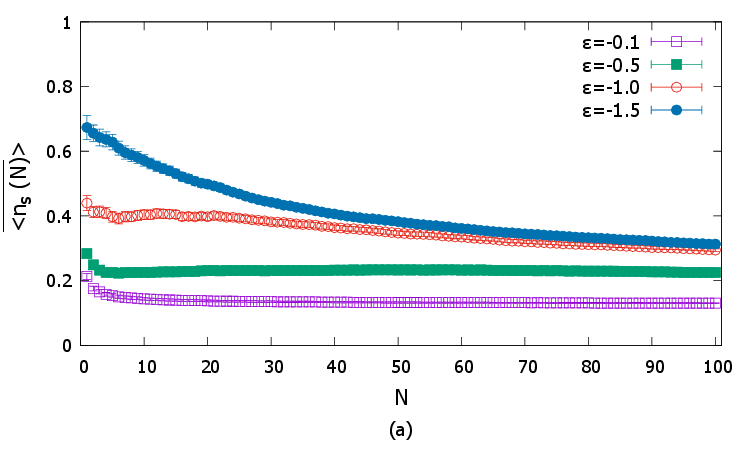}
\includegraphics[width=8.2cm]{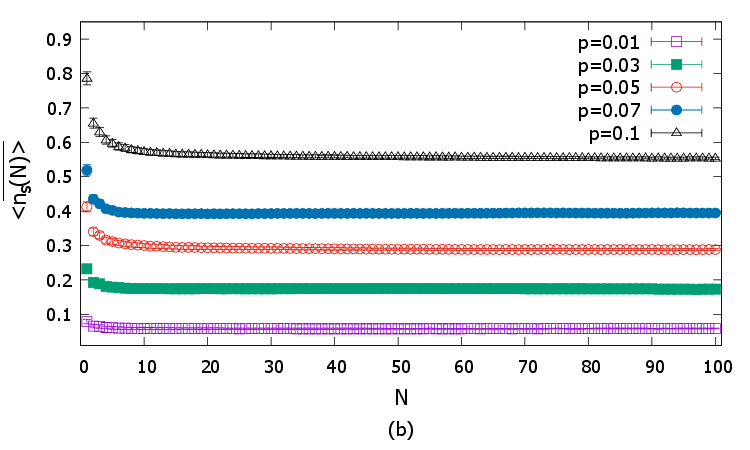}
\includegraphics[width=8.2cm]{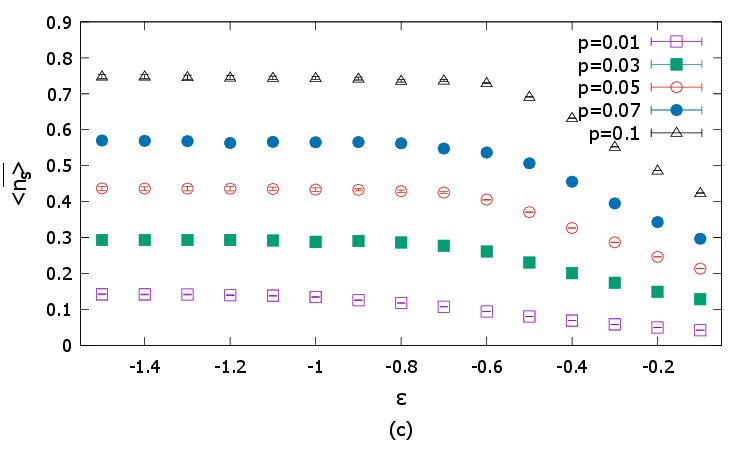}
\includegraphics[width=8.2cm]{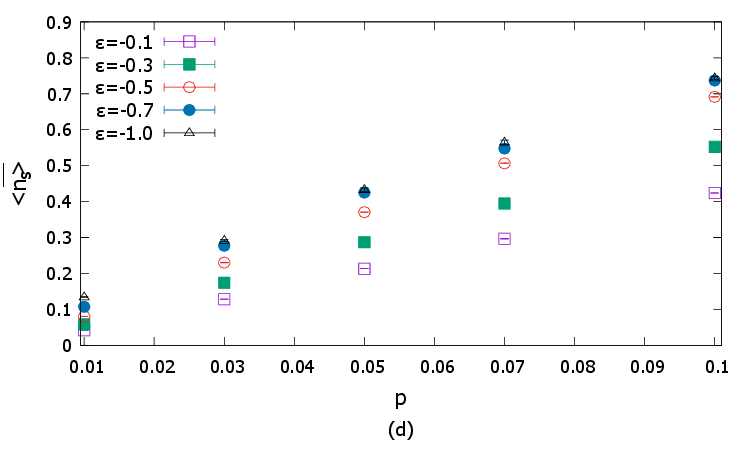}
\end{center}
{\caption{\label{ns} The averaged number of adsorbed obstacles per one monomer as given by $\langle {\overline{n_{s}(N)}} \rangle $  as function of $N$ a) at fixed $p=0.03$ and several values of  $\varepsilon$ and b) at fixed $\varepsilon=0.3$ and several values of  $p$.  Asymptotic values of  $\langle {\overline {n_{s}}} \rangle$, obtained on the basis of fitting function (\ref{bs}), as functions of $\varepsilon$ c) and function of $p$ d).
}}
\end{figure}

{ We considered also the behaviour of the total number of contacts $N_s(N)$ between the monomers of SAW and adsorbed obstacles, as defined in Eq.~(\ref{Nsdef}). The reduced value $n_s(N)=N_s(N)/N$ gives thus the averaged number of adsorbed obstacles per one monomer. The dependencies of this parameter on $N$, $\varepsilon$, and $p$ are presented in Fig. \ref{ns}. Note that the probabilities $g(i,p)$ for a single monomer of SAW to encounter the number of obstacles, $i$ in its neighborhood on a simple cubic lattice  with concentration $p$ of obstacles can be roughly estimated as: $g(i,p)=C(4,n)p^i(1-p)^{4-i}$ with $C(4,n)$ being binomial coefficient. The multiple adsorption of obstacles on a monomer ($i>1$) is thus very unlikely for small values of $p$. Indeed, $\langle n_s \rangle$ attains very small values (below 1) in the range of $p$ and $\varepsilon$ values under consideration (as presented in Fig. \ref{ns}), while it is growing both with increasing the obstacle concentration and of binding strength.}

{We can compare obtained range of $n_s(N)$ with typical experimental values provided by Tyagi et al. \cite{Tyagi2020} for the EDTA-containing polymeric chelants and ions of copper. This data is provided in Fig.4b in their study and is termed as ``chelation capacity''. The values between $0.3$ and $0.5$ are reported depending on $pH$. This interval of values for $n_s(N)$ is found in Fig.~\ref{ns}c at $0.03<p<0.06$ in the asymptotic limit of high binding. The condition $pH=7$ is characterized by fully deprotonated state of EDTA and, hence, the strongest chelation ability \cite{latva1996, licup2024}. Therefore, it can be associated with the larger values for $|\varepsilon|$ in a current model. With the increase of $|\varepsilon|$ from $0$ to $0.6$, $n_s(N)$ raises as well, mimicking increase of the chelation capacity with the growth of $pH$ from $3$ to $7$, as reported in Ref.~\cite{Tyagi2020}.}

Dependence of $\langle \overline{n_{{\rm bond}}(N)} \rangle$ on $N$ at fixed $\varepsilon=-0.3$ and several values of obstacle concentration $p$ is given in Fig.~\ref{nbond}b. Here we notice the non-linear dependence of $\langle \overline{n_{{\rm bond}}(N)} \rangle$ on $p$, in contrast to $\langle \overline{n_a(N)} \rangle$ in Fig. \ref{Na}b-c. The curves for $\langle \overline{n_{{\rm bond}}(N)} \rangle$, shown in this figure,  come very close at $p\geq 0.05$, indicating no further raise of adsorption strength when obstacles concentration increases beyond $0.05$.
Adsorption efficiency $\langle \overline{n_{{\rm bond}}(N)} \rangle$ gradually reaches its asymptotic value of $\langle \overline{n_{{\rm bond}}} \rangle$, which can be estimated again based on finite-size scaling law
\begin{equation}
\langle \overline{n_{{\rm bond}}(N)} \rangle = 
 \langle \overline{ n_{{\rm bond}}} \rangle+B/N. \label{bs}
\end{equation}
Our estimates for the asymptotic values for $\langle \overline{n_{{\rm bond}} }\rangle$, as functions of $\varepsilon$ at various $p$, are given in Fig. \ref{nbond}c. These show good level of saturation at $p\geq 0.03$.

\begin{figure}[b!]
 \begin{center}
\includegraphics[width=8.2cm]{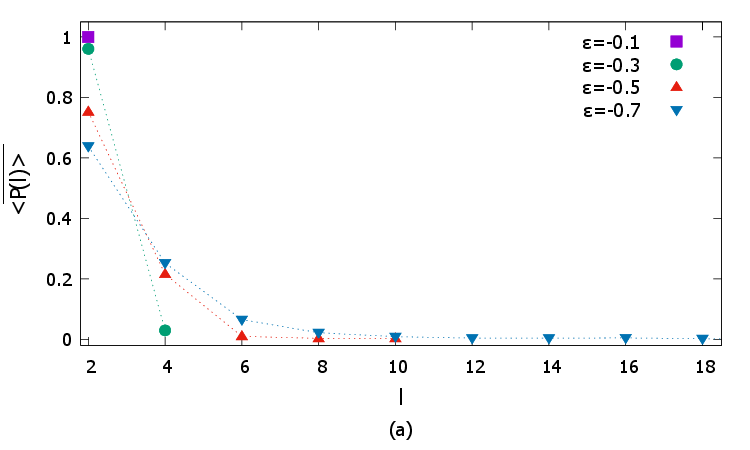}
\includegraphics[width=8.2cm]{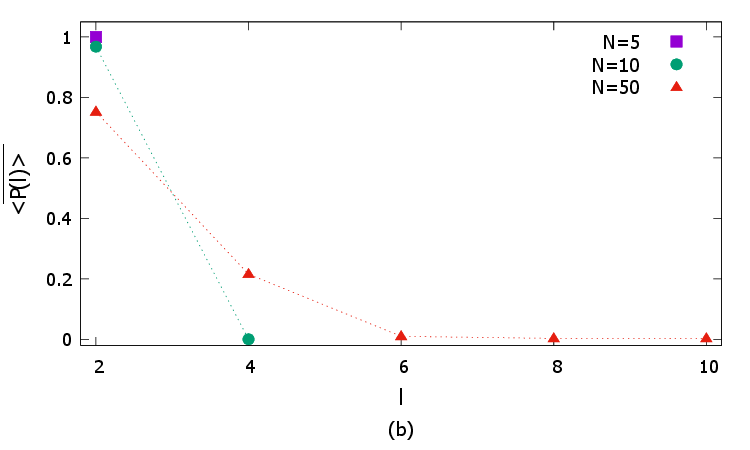}
\end{center}
\caption{\label{probloop} Distribution of lengths $l$ of ``loops'' inside the polymer chain established by contacts of monomers with obstacles at $p=0.03$. a) Fixed chain length $N=50$ and several values of $\varepsilon$; b) fixed $\varepsilon=-0.5$ and several values of $N$. Lines are guides to the eye. }
\end{figure}

{ As follows from the discussion above, the differences in adsorption efficiency and strength demonstrated by short and long chains can be explained by different set of conformations that are accessible to them. In particular, longer chains are more prone to form highly coiled conformations, as well as ``loops'' of $l$ monomers that appear between $(m)$th and $(m+l)$th monomer when both form a bond with the same obstacle. The presence of such loops would explain higher values of $\langle \overline{n_{{\rm bond}}(N)} \rangle$ for larger $N$, see Fig. \ref{nbond}a-c.}

{ The case $l=0$ corresponds to a single contact of a polymer chain with an obstacle and is trivial. Therefore, we will be interested in not-trivial loops, $l>0$, only. An example of the distribution of such loop lengths at fixed $N=50$ is given in Fig. \ref{probloop}a. For the case of a small binding energy magnitude, $\varepsilon=-0.1$, the most probable loops are rather short: $l=2$ or $4$, indicating that the same obstacle can be adsorbed only by the odd or even neighbours along the chain. Thus, the conformation of the polymer coil and its effective dimensions, are not affected by the presence of obstacles. However, with the increase $|\varepsilon|$,} larger values of $l$ are observed (yet with considerably smaller probability): polymer chain thus attains a slightly more compact conformation in this case, caused by establishing bridges between monomers with a considerable mutual separation along the chain. In Fig. \ref{probloop}b, the loop distributions at fixed  $\varepsilon=-0.5$ and a range of chain lengths $N$ are presented. For the very short chains ($N=5$), only very short loops of $l=2$ can be found, while the probability of larger loops grows with increasing the total length of a polymer chain.


\begin{figure}[t!]
 \begin{center}
\includegraphics[width=8.2cm]{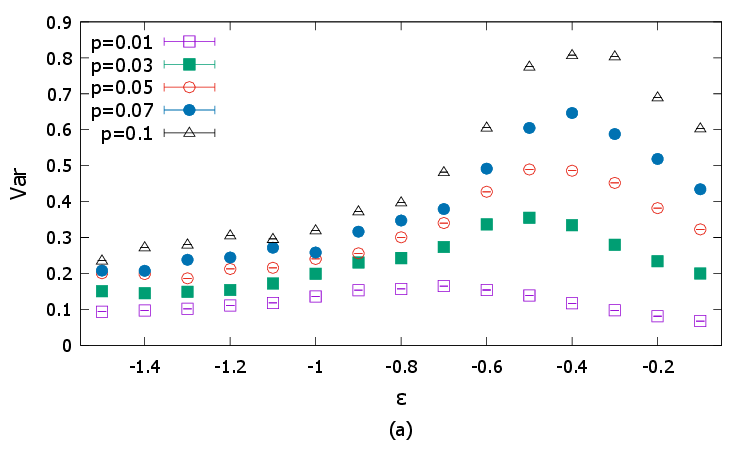}
\includegraphics[width=8.2cm]{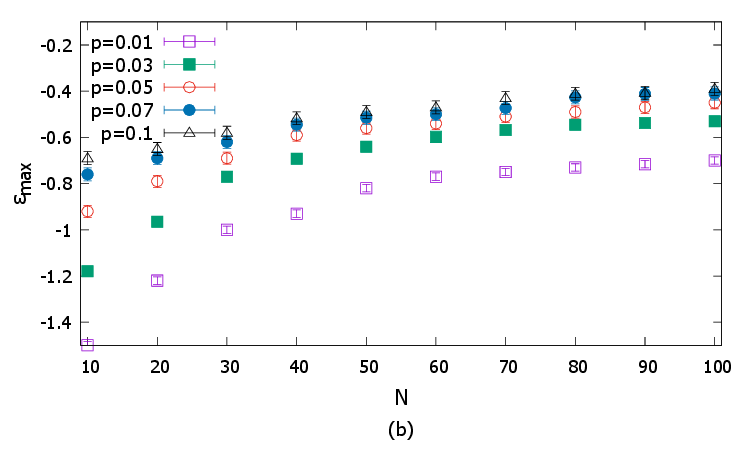}
\end{center}
\caption{\label{Cv} a) Variance of the total number of monomer-obstacle contacts, as defined by Eq. (\ref{Var}) as function of binding energy $\varepsilon$ at $N=100$. b) The peak values $\varepsilon_{\mathrm{max}}$ at various $p$ as functions of $N$. }
\end{figure}

To examine a role played by the binding energy $\varepsilon$, we also examined the fluctuations of the combined characteristic, the product of the adsorption efficiency and adsorption strength, given by the total number of chelation contacts $N_s(N)$. It is measured by evaluation of the variance:
\begin{equation}
\it{Var}=\frac{1}{N}(\langle \overline{N^2_s(N)} \rangle - \langle \overline{N_s} \rangle^2). \label{Var}
\end{equation}
Our results are presented in Fig. \ref{Cv}a, at fixed polymer length $N=100$. The presence of the peak in this plot at characteristic binding energy $\varepsilon_{\mathrm{max}}$, indicates the maximum fluctuation strength, and may signal the presence of a transition or crossover between physically different states.  Note, that $\varepsilon_{\mathrm{max}}$ increases with $N$, as indicated in Fig. \ref{Cv}b.  The estimates for the peak positions may be interpreted as the critical values of $\varepsilon$ (adsorption point), separating the ``free'' ($|\varepsilon|<|\varepsilon_{\mathrm{max}}|$) and ``adsorbed'' ($\varepsilon=\varepsilon_{\mathrm{max}}$) states of a polymer. In the latter case, the polymer reaches its maximal possible efficiency in establishing bridging bonds with obstacles (cf. Fig. \ref{Na}b), and further increase of the binding energy magnitude $|\varepsilon|$ beyond $|\varepsilon_{\mathrm{max}}|$ has no effect. This situation can be related to a standard problem of polymer adsorption on a (fractal) surface \cite{Janke12}.

\begin{figure}[b!]
 \begin{center}
\includegraphics[width=8.2cm]{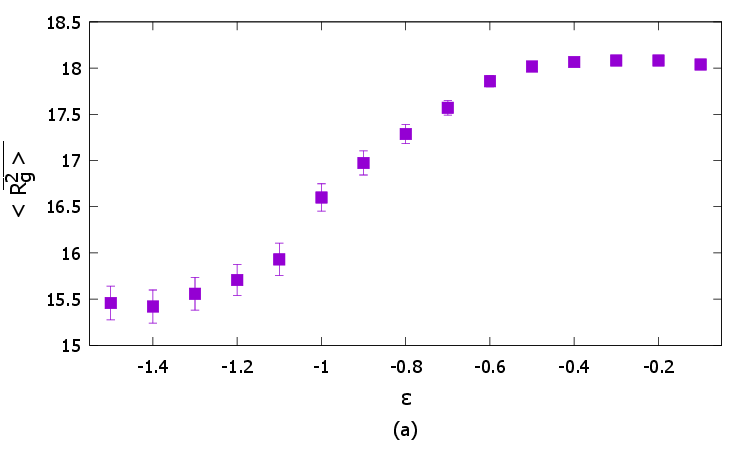}
\includegraphics[width=8.2cm]{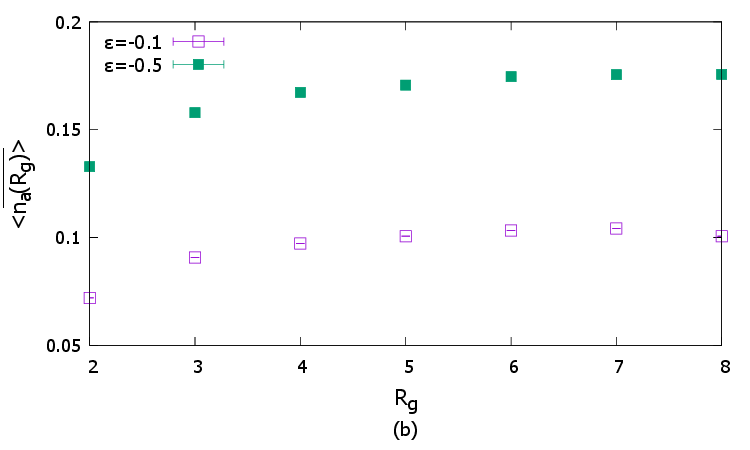}
\end{center}
\caption{\label{Rgpsilon} a) Average mean-squared gyration radius ${\overline {\langle R_g^2 \rangle }}$ of SAW trajectory as function of binding energy $\varepsilon$ at $N=50$, $p=0.03$. b) Correlation between the values of gyration radius $R_g$ of SAW trajectory with $N=50$ and the adsorbtion efficiency at $p=0.03$. }
\end{figure} 

As a size characteristic of typical conformation of  polymer chain of length $N$ in a presence of obstacles, we analyzed the gyration $R_g$, defined by:
\begin{equation}
R_g^2=\frac{1}{N} \sum_{i=1}^N (\vec{R}_i-\vec{R}_{CM})^2,
\end{equation}
where $\vec{R}_i$ provides position of $i$th monomer, and $\vec{R}_{CM}$ is the center of mass of a chain. Our data for $\langle \overline{ R_g^2} \rangle$, as the function of $\varepsilon$ at fixed $N=100$ and the obstacles concentration of $p=0.03$, are given in Fig. \ref{Rgpsilon}a. In the case of a zero binding energy, $\varepsilon=0$, the presence of obstacles causes purely steric effect on the conformational properties of the {SAW}, {promoting more extended} polymer configuration {(indicated by higher values of $\langle \overline{ R_g^2} \rangle$) as compared with the no obstacles case} \cite{Klimov1992}; this effect becomes more pronounced with the increase of the concentration of obstacles $p$. For the non-zero binding energy, $\varepsilon<0$, {the SAW conformations are also affected, but now by the attraction of its monomers to obstacles, and the effect is an opposite: the effective size of SAW gradually decreases} with the increase of $|\varepsilon|$. Similar behaviour of $R_g$ was observed, in particular, in Monte Carlo simulations performed in Ref. \cite{Zherenkova1998}.  Note that such ``compactification" of a polymer chain with the increase of $|\varepsilon|$ can be explained by the increase of the lengths of ``loops'' within the macromolecule in this case, as we discussed earlier (see Fig. \ref{probloop}). { Since the SAW is an irregular structure with its effective size $R_g$ taking on a range of different values in an ensemble of possible SAW configurations, we analyzed the correlation between values of $R_g$ and the adsorption efficiency $\overline{\langle n_a(R_g)} \rangle$, observed in performing averaging only on the configurations with given  $R_g$.  Corresponding results are presented in Fig. \ref{Rgpsilon}b. The smaller values of $R_{g}$ correspond to more compact SAW configurations with higher fraction of ``inner'' monomers, screened from contacts with obstacles. Indeed, the values of $\overline{\langle n_a(R_g)} \rangle$ are smaller for the case of more compact SAW and increase gradually with an increase of  $R_g$. 
}

Therefore, our analysis of the adsorption efficiency, $\langle \overline{n_a(N)}\rangle$, and adsorption strength, $\langle \overline{n_{{\rm bond}}(N)}\rangle$, in the short and long chain regimes, respectively, focuses on the following two factors.

The \textit{first factor} is related to the ability of a polymer chain to form a coiled conformation. In a regime of a short chain, $N<10$, the chain is too short to coil and exhibits predominantly stretched conformations. In this case, the chain adsorbs larger amount, $\langle \overline{n_a(N)}\rangle$, of obstacles, but with a small number of contacts, $\langle \overline{n_{{\rm bond}}(N)}\rangle$, in a course of its growth, see Figs.~\ref{Na}a and \ref{nbond}a. Adsorption efficiency, $\langle \overline{n_a(N)}\rangle$, in this case is strongly affected by the binding energy $\varepsilon$. In contrary, in the regime of a long chain, $N>10$, due to entropic factor, the chain forms a coil, which is a much more compact object. This increases the probability of a newly grown segments of a chain to meet the same obstacles, that have been already absorbed on the earlier stages of the chain growth. In this case, $\langle \overline{n_a(N)}\rangle$ will be lower, whereas $\langle \overline{n_{{\rm bond}}(N)}\rangle$ will be higher than in the regime of a short chain, as evidenced by Figs.~\ref{Na}a and \ref{nbond}a. Hence, adsorption efficiency here is predominantly restricted by the relatively compact conformations of coiled chains and, therefore, displays a very weak or no dependence on the binding energy $\varepsilon$. It is interesting to note, that the SAW enters asymptotic scaling regime at certain minimum length, indeed, and the estimates made by the dissipative particle dynamics simulations provide the value of about $N\approx 10$ \cite{Ilnytskyi2007, Kalyuzhnyi2016}. The ``bridging loop'' statistics, as discussed earlier (Fig. \ref{probloop}), also clarifies the difference between the short and long chain regimes as indicated by $\langle \overline{n_a(N)} \rangle$ and $\langle \overline{n_{{\rm bond}}(N)}\rangle$ behaviour. Indeed,  establishing of longer loops at larger $N$ prevents the monomers inside of them to ``find'' another obstacles and explains increasing of the average number of bonds per obstacle with increasing $N$ (cf. Fig. \ref{nbond}a).

The \textit{second factor} is related to the total functionality of the monomers of a polymer chain with respect to formation of bridges. In this context, let us estimate the number of the nearest neighbour sites for each monomer of a chain, that are not occupied by a polymer chain itself. In a general case of the $d$-dimensional regular lattice, the first and the last monomers of a chain has $m_\mathrm{end}=2d-1$ nearest neighbours, whereas for the internal monomers of a chain this number is $m_\mathrm{inside}=2d-2$ (note that we neglect the self-avoidance effect in this estimate). At a given concentration of obstacles $p$, the probability of $N$-monomer chain to encounter an ion per one monomer can be thus estimated as
\begin{eqnarray}
&&{\rm prob}\simeq p(2m_{end}+(N-2)m_\mathrm{inside})/N=\nonumber\\
&&=2p(d-1+1/N). 
\end{eqnarray}
The only dependence of this probability on the chain lengths $N$ is contained in the term proportional to $1/N$, which decays with the increase of a polymer length $N$. Both factors considered above, contribute towards the increase of adsorbing efficiency of shorter chains as compared to their longer counterparts, at least within the realms of the model being suggested.

\section{Conclusions}

Mixing polymers with organic or inorganic particles opens up formation of new composite systems, such as colloidal suspensions, microemulsions, micellar solutions, etc. Depending on the relation between the particle diameter and characteristic dimensions of a polymer, different scenarios are observed. For relatively large particles, polymers may act as connecting spacers, coating medium, induce depletion based flocculation, promote necklace-like aggregates, etc. For relatively large polymers, one encounters
adsorption of dissolved polymer on a small sized colloid particles. In many of theoretical studies of such system, linear polymer is represented via a SAW on a lattice, neglecting atomic scale chemical details and concentrating instead on a large scale conformation statistics.

In the extreme case of a very small particles, one can relate such a system to the problem of chelation of heavy metal ions by polymeric chelants, which is of a paramount importance from both biomedical aspect related to deintoxication and environmental perspective of wastewater cleaning. The technique occupies a broad range of scales, and while atomistic simulations focus on the details of formation of chelation complexes, the concept of a SAW on a obstacle-ridden lattice emphasizes the role of polymer conformations in chelation process on a large scale. This is the topic of a current study.

We consider a single polymer chain of the length $N$ as a SAW on the $3D$ lattice, where the latter contains a set of randomly distributed obstacles with the concentration $p$. The SAW is grown at $1000$ different realizations of quenched disorder enabling to mimic diffusion of obstacles implicitly. The PERM Monte Carlo simulations algorithm is applied and the behaviour of the system is analyzed at a range of $N$, $p$, and the polymer-obstacle binding energy $\varepsilon$ (measured in units of $k_BT$). For each realization of disorder, $10^6$ PERM iterations for a SAW is performed, and $10$ chains on average are generated per each iteration.

The study focuses on the following set of characteristics. Adsorption efficiency, $\langle \overline{n_a(N)} \rangle$, is associated with the average number of adsorbed obstacles per one monomer, whereas its strength, $\langle \overline{n_{{\rm bond}}(N)}\rangle$, is characterized by the average number of polymer-obstacle bonds,  per each obstacle. Obstacle-induced formation of polymer loops is quantified by the distribution $P(l)$ of the loop sizes $l$. The variance, $\it{Var}$, of the total number of polymer-obstacle bonds in the system, as well as the average shape of a SAW, in terms of its radius of gyration, $R_g$, are also considered. All these properties are examined for a wide range of $N$, $p$ and $\varepsilon$.

We found that the polymer length, $N$, plays a crucial role in defining the characteristics of adsorption. Namely, the short chains regime, $N<10$, is characterized by high adsorption efficiency, $\langle n_a(N)\rangle$, but low adsorption strength, $\langle \overline{n_{{\rm bond}}(N)}\rangle$. Both are found to depend strongly on the binding energy $\varepsilon$. Predominant role played by binding energy in this case is explained by a relatively weak ability of a short chain to form a coil. As the result, it just makes its relatively straight pass through the lattice being driven strongly by adsorption on the nearest obstacles on its way. {Higher adsorption efficiency of short chains is partly supported by existing experiments on sedimentation of silica colloids \cite{Gonzlez2011}, but the lattice model in our study is insufficient to take into account all relevant factors.}

In contrast to this, the long chain regime, $N>10$, is characterized by lower adsorption efficiency, $\langle \overline{n_a(N)} \rangle$, and higher adsorption strength, $\langle \overline{ n_{{\rm bond}}(N)} \rangle$, as well as a weak dependence on the binding energy $\varepsilon$. This is explained by the ability of longer chains to form compact coils, in which case a chain more readily creates new contacts with already adsorbed obstacles within a coil interior than engages new obstacles outside of it. This scenario is further supported by an increased probability of a chain to form loops, as evidenced by the nonzero values of $P(l)$ at $l>2$. Adsorption in this case is principally dominated by conformation properties of a polymer chain and not by the binding energy $\varepsilon$.

Analysis of both $\it{Var}$ and $R_g$ for a long SAW of $N=100$ monomers indicates the presence of a specific, ``critical", value for the binding energy: $\varepsilon_c\approx -0.4$. The variance $\it{Var}$ at $\varepsilon_c$ exhibits a maximum, which grows with the increase of the concentration of obstacles, $p$, whereas $R_g$ decays at $|\varepsilon|>|\varepsilon_c|$, reminding the behaviour of the susceptibility and magnetization, respectively, during magnetic phase transitions. This indicates the presence of some transition or crossover effects at $\varepsilon=\varepsilon_c$ related to adsorption, but currently we can not suggest more definite interpretation.

Application and direct interpretation of our findings in terms of the efficiency of real polymer chelants is not straightforward, though. Firstly, the volume packing of monomers and obstacles in the lattice SAW is restricted by imposed spatial discretization of a lattice. E.g., higher adsorbing abilities of the chain ends of a lattice-based SAW, as discussed in the study, may not be exactly the case for real polymers, where both chain ends are typically functionalized by chelation-neutral end groups. Secondly, the interpretation of a monomer-obstacle bond is nontrivial, as on the scale of a polymer on a lattice it rather represents a whole chelation complex and not an individual chelation bond. In this case, the second monomer-obstacle bond (and, hence, a loop) can only be formed if the existing bond represents unsaturated chelation complex involving only 2-3 chelating groups of a real polymer chelant. One might need to introduce additional variable of the bond type (saturated vs unsaturated) to account for such statistics. Finally, we consider a single polymer chain case, whereas one would normally use some non-zero concentration of polymer chelants. Therefore, the question of efficiency between a single long chain (with possible loop formation) and of a set of short chains of equivalent molecular weight (with their ability to form ion-aided interchain bridges \cite{Rosthauser1981, Johann2019}), is to be considered, too. 

Given the known limitations of a model approach, the results {obtained provide reasonable coarse-grained description of chelation of metal ions by polymeric chelants. Some new insights are found concerning the role of conformations statistics in the process of chelation within a wide intervals of the chelant lengths, binding energies, and ions concentrations. In particular, the values for the chelation capacity (average number of adsorbed particles per monomer) are within the interval of typical experimental values \cite{Tyagi2020}, as well as the dependence of this characteristic on $pH$ (directly connected to the protonation level of a chelant) is related to the variation of a magnitude $|\varepsilon|$.} These findings, complemented by incorporation of relevant chemical details, may provide a basis of a future, multi-scale approach, that is able to predict chelation efficiency of particular polymeric chelants. Strong dependence of the adsorption scenario on the polymer length, polymer coil structure, related excluded volume effects and adsorption strength, suggest performing following studies involving more complex, branched, polymeric architectures.

\section*{Acknowledgements}

The work was supported by Academy of Finland, reference number 334244.

\bibliography{Chelation}

\end{document}